\documentclass{PoS}
\usepackage{graphicx}
\usepackage{amsmath}
\usepackage{lineno}
\input{babarsym.tex}

\title{Recent Results on $T$ and $CP$ Violation at {\babar}}

\ShortTitle{Recent Results on $T$ and $CP$ Violation at {\babar}}

\author{\speaker{Alejandro P\'erez P\'erez}\\
        INFN - sezione di Pisa\\
        E-mail: \email{luis.alejandro.perez@pi.infn.it}\\
        {\it On behalf of the {\babar} Collaboration}}

\abstract{$CP$-violation ($CPV$) and Time-reversal violation ($TRV$) are intimately related through the $CPT$
theorem: if one of these discrete symmetries is violated the other one has to be violated in such a way to 
conserve $CPT$. Although $CPV$ in the $B^0\bar{B}^0$ system has been established by the B-factories, implying 
indirectly TRV, there is still no direct evidence of $TRV$. We report on the observation of $TRV$ in the 
B-meson system performed with a dataset of $468\times10^6$ $B\bar{B}$ pairs produced in $\Upsilon(4S)$ decays 
collected by the {\babar} detector at the PEP-II asymmetric-energy $e^+e^-$ collider at the SLAC National 
Accelerator Laboratory. We also report on other $CPV$ measurements recently performed on the B-meson system.}

\FullConference{XXI International Workshop on Deep-Inelastic Scattering and Related Subject -DIS2013,\\
		22-26 April 2013\\
		Marseilles,France}

\begin{document}
\section{First direct observation of $T$-reversal violation in B-mesons}

The Cabibbo-Kobayashi-Maskawa ($CKM$) matrix mechanism~\cite{CKM} for the quark mixing describes all
transitions between quarks in terms of only four parameters: three rotation angles and one irreducible 
phase. This irreducible phase being the only source of $CPV$ in the standard model (SM). $CPV$ has 
been well established both in the K-meson~\cite{CPV_K} and B-meson~\cite{CPV_B} systems, being consistent 
with the $CKM$ mechanism. Local Lorentz invariant quantum field theories imply $CPT$ invariance~\cite{CPT}, 
in agreement with all experimental evidence up to date~\cite{CPT_exp}. It is therefore expected that 
the $CP$-violating weak interaction also violates $T$-reversal.

In stable systems, a signature of $TRV$ would be a non-zero expectation value for a $T$-odd observable, 
e.g. neutron or electron electric dipole moments, but no such observation has been made up to date. 
The only evidence of $TRV$ has been found in the neutral K-meson system, with the measurement of 
the difference between the probabilities of $K^0 \rightarrow \bar{K}^0$ and $\bar{K}^0 \rightarrow K^0$ 
transitions for a given elapsed time~\cite{TRV_K}. However, since this flavour mixing asymmetry violates both 
$CP$ and $T$, it is impossible to disentangle $TRV$ from $CPV$. In unstable systems, $TRV$ can be explored 
by studying a process under the $t\rightarrow -t$ transition combined with the exchange of $|in\big>$ and 
$|out\big>$ states, which can be experimentally challenging to achieve. As an example, comparing the rates 
of $B^0 \rightarrow K^+\pi^-$ and $K^+\pi^− \rightarrow B^0$ is not feasible due to the need to prepare the 
initial state and to disentangle weak from strong effects. However, the coherent production of B-mesons pairs 
at the B-factories, offers a unique opportunity to compare couple of processes where the initial and final 
states are exchanged by Time-reversal.

The experimental method described in Ref.~\cite{TRV_method} proposes to use the entangled quantum state $|i\big>$ 
of the two neutral B-mesons produced through the $\Upsilon(4S)$ decay. This two-body state usually written in terms 
of the flavour eigenstates, $B^0$ and $\bar{B}^0$, can be as well expressed in terms of mutually 
orthogonal $B_+$ and $B_-$ $CP$-eigenstates, which decay to $CP=+1$ and $CP=-1$, respectively: 
$|i\big> = \frac{1}{\sqrt{2}} [B^0(t_1)\bar{B}^0(t_2) - \bar{B}^0(t_1)B^0(t_2)] = \frac{1}{\sqrt{2}} [B_+(t_1)B_-(t_2) - B_-(t_1)B_+(t_2)]$.
Experimentally, the $B_+$ and $B_-$ states are defined as the neutral B states filtered by the decay to 
$CP$ eigenstates $J/\psi K^0_L$ ($CP=+1$) and $J/\psi K^0_S(\rightarrow\pi\pi)$ ($CP=-1$). We define reference 
transitions and their $T$-transformed counterparts (see table~\ref{tab:TRV_transitions}) and compare their 
transition rates as a test for $T$-reversal. The notation $(X,Y)$ denotes the final states of the time ordered 
B-meson decays from the entangled state, with $B\rightarrow X$ ($B\rightarrow Y$) the earlier (later) decay. 
The time difference between the decays, $\Delta t = t_Y - t_X$, is then positive by definition. As an 
illustration, the pair of final states $(\ell^+,~J/\psi K^0_S)$ denotes a $B^0\rightarrow \ell^+ X$ decay,
meaning that at that time the other $B$ in the event is a $\bar{B}^0$, followed in time by a 
$B\rightarrow J/\psi K^0_S$ decay, projecting to a $B_-$. The full process is the transition $\bar{B}^0\rightarrow B_-$. 
A difference between this rate $\bar{B}^0\rightarrow B_-$ and its $T$-transformed one is an indication of 
$TRV$. As shown in table~\ref{tab:TRV_transitions}, a total of four $T$-reversed transitions can be studied. 
The experimental analysis exploits identical reconstruction algorithms and 
selection criteria of the {\babar} time-dependent $CP$ asymmetry measurement in $B \rightarrow c\bar{c} K^{(*)0}$ 
decays~\cite{CPV_B}. The {\it flavor tagging} is combined for the first time with the {\it CP tagging}, as required for the 
construction of $T$-transformed processes.

\begin{table}[hbt!]
\begin{center}
\begin{TableSize}
\begin{tabular}{cc}
\hline
Reference transition $(X,Y)$  & T-transformed transition $(X,Y)$ \\
\hline
$B^0\rightarrow B_+ (\ell^-,J/\psi K^0_L)$ & $B_+\rightarrow B^0 (J/\psi K^0_S,\ell^+)$ \\
$B^0\rightarrow B_- (\ell^-,J/\psi K^0_S)$ & $B_-\rightarrow B^0 (J/\psi K^0_L,\ell^+)$ \\
$\bar{B}^0\rightarrow B_+ (\ell^+,J/\psi K^0_L)$ & $B_+\rightarrow \bar{B}^0 (J/\psi K^0_S,\ell^-)$ \\
$\bar{B}^0\rightarrow B_- (\ell^+,J/\psi K^0_S)$ & $B_-\rightarrow \bar{B}^0 (J/\psi K^0_L,\ell^-)$ \\
\hline
\end{tabular}
\end{TableSize}
\end{center}
\caption{\em Reference transitions and their T-transformed.
\label{tab:TRV_transitions}
}
\end{table}

The decay rate is proportional to
$g^{\pm}_{i,j}(\Delta t) \propto e^{-\Gamma_d\Delta t}\{1 + S^{\pm}_{i,j} \sin(\Delta m_d\Delta t) + C^{\pm}_{i,j} \cos(\Delta m_d\Delta t)\}$,
where $i$ denotes $B^0$ or $\bar{B}^0$, $j$ denotes $J/\psi K^0_S$ or $J/\psi K^0_L$, and $\pm$ indicates 
whether the flavour final state occurs before ($+$) or after ($-$) the $CP$ decay. $\Gamma_d$ is the average decay 
width, $\Delta m_d$ is the $B^0\bar{B}^0$ mass difference. There are eight distinct sets of $C^\pm_{i,j}$ 
and $S^\pm_{i,j}$ parameters. An unbinned maximum likelihood fit is performed to the $B^0$, $\bar{B}^0$, 
$c\bar{c}K^0_S$ and $J/\psi K^0_L$ samples, to extract the $C^\pm_{i,j}$ and $S^\pm_{i,j}$ 
parameters. Out of this set of fitted parameters, a different set of $T$, $CP$ and $CPT$ violation parameters
can be built, $\Delta C^{\pm}_i$, $\Delta S^{\pm}_i$ (with $i = T, CP, CPT$) which are constructed as differences 
of the $C^\pm_{i,j}$ and $S^\pm_{i,j}$ for symmetry-transformed transitions (see 
table~\ref{tab:TRV_parameters_results}). Any deviation of the $(\Delta C^{\pm}_i,\Delta S^{\pm}_i)$ from 
$(0,0)$ signals the violation of the corresponding symmetry.

\begin{figure}[hb!]
\begin{center}
\includegraphics[width=4.5cm,keepaspectratio]{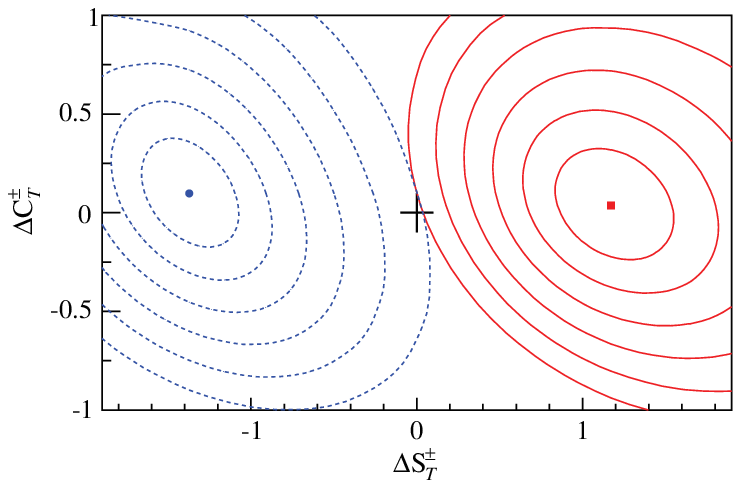}
\includegraphics[width=4.5cm,keepaspectratio]{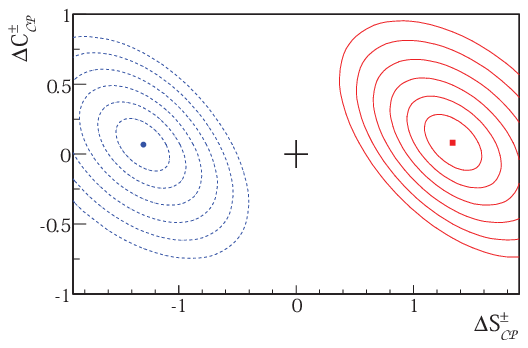}
\includegraphics[width=4.5cm,keepaspectratio]{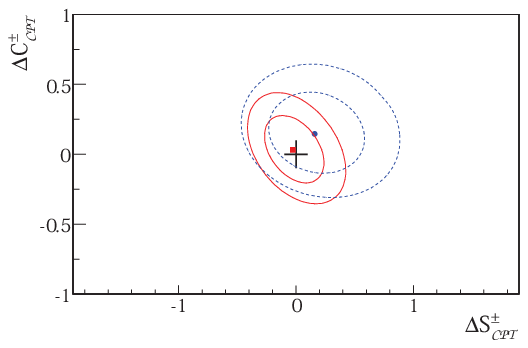}
\end{center}
\caption
{\label{fig:TRV_results}
{\em Confidence level contours at intervals of $1\sigma$ for $T$- (left), $CP$- (middle) and $CPT$- (right) 
differences results. $\Delta S^+_i$ and $C^+_i$ ($\Delta S^-_i$ and $C^-_i$) are shown as a blue 
dashed (solid red) curves. The no-violation point of the corresponding symmetry is indicated with a 
cross (+).
}}
\end{figure}

\begin{table}[hbt!]
\begin{center}
\begin{TableSize}
\begin{tabular}{cc|cc}
\hline
Parameter & Measurement & Parameter & Measurement\\
\hline
$\Delta S^+_T = S^-_{\ell^-,K^0_L} - S^+_{\ell^+,K^0_S}$ & $-1.37 \pm 0.14 \pm 0.06$ & 
$\Delta C^+_T = C^-_{\ell^-,K^0_L} - C^+_{\ell^+,K^0_S}$ & $0.10 \pm 0.14 \pm 0.08$  \\
$\Delta S^-_T = S^+_{\ell^-,K^0_L} - S^-_{\ell^+,K^0_S}$ & $1.17 \pm 0.18 \pm 0.11$ & 
$\Delta C^-_T = C^+_{\ell^-,K^0_L} - C^-_{\ell^+,K^0_S}$ & $0.04 \pm 0.14 \pm 0.08$  \\

$\Delta S^+_{CP} = S^+_{\ell^-,K^0_S} - S^+_{\ell^-,K^0_S}$ & $-1.30 \pm 0.11 \pm 0.07$ & 
$\Delta C^+_{CP} = C^+_{\ell^-,K^0_S} - C^+_{\ell^-,K^0_S}$ & $0.07 \pm 0.09 \pm 0.03$  \\
$\Delta S^-_{CP} = S^-_{\ell^-,K^0_S} - S^-_{\ell^+,K^0_S}$ & $1.33 \pm 0.12 \pm 0.06$ & 
$\Delta C^-_{CP} = C^-_{\ell^-,K^0_S} - C^-_{\ell^+,K^0_S}$ & $0.08 \pm 0.10 \pm 0.04$  \\

$\Delta S^+_{CPT} = S^-_{\ell^+,K^0_L} - S^+_{\ell^+,K^0_S}$ & $-1.30 \pm 0.11 \pm 0.07$ & 
$\Delta C^+_{CPT} = C^-_{\ell^+,K^0_L} - C^+_{\ell^+,K^0_S}$ & $0.07 \pm 0.09 \pm 0.03$  \\
$\Delta S^-_{CPT} = S^+_{\ell^+,K^0_L} - S^-_{\ell^+,K^0_S}$ & $1.33 \pm 0.12 \pm 0.06$ & 
$\Delta C^-_{CPT} = C^+_{\ell^+,K^0_L} - C^-_{\ell^+,K^0_S}$ & $0.08 \pm 0.10 \pm 0.04$  \\
\hline
\end{tabular}
\end{TableSize}
\end{center}
\caption{\em Measured values of the $T$, $CP$ and $CPT$ difference parameters. The first 
uncertainty is statistical and the second systematic. The indexes $\ell^{\pm}$ and 
$K^0_S$/$K^0_L$ are described in the text.
\label{tab:TRV_parameters_results}
}
\end{table}

The results on the $T$, $CP$ and $CPT$ asymmetries are shown in table~\ref{tab:TRV_parameters_results}. The 
significance of the corresponding differences is shown graphically in figure~\ref{fig:TRV_results}, with the 
two-dimensional contours in the $(\Delta S^{\pm}_i,\Delta C^{\pm}_i)$ planes ($i=T,CP,CPT$). time-reversal 
violation is clearly established, with the exclusion of the $(0,0)$ point with a 
significance of $14\sigma$. $CP$-violation is also observed at the level of $16\sigma$. No evidence 
of $CPT$-violation is observed, the measurement being consistent with the conservation hypothesis 
within the $1\sigma$ level~\cite{TRV_paper}.

\section{$CP$-violation in $B^0\bar{B}^0$ mixing}

Two of the three types of $CP$-violation that can be observed in neutral B-mesons systems have been well established, 
i.e. $CP$-violation in direct $B^0$ decays and in the interference between mixing and decay~\cite{CPV_B}. The third 
one, $CP$-violation in mixing has so far eluded observation. The weak-Hamiltonian eigenstates are related to the 
flavour eigenstates as $|B_{L,H}\big> = p|B^0\big> \pm q|\bar{B}^0\big>$. The asymmetry between the oscillation 
probabilities $P = P(B^0\rightarrow \bar{B}^0)$ and $\bar{P} = P(\bar{B}^0\rightarrow B^0)$ is defined as: 
$A_{CP} = \frac{\bar{P}-P}{\bar{P}+P} = \frac{1 - |q/p|^4}{1 + |q/p|^4}\simeq 2(1-|q/p|^2)$. Hence, there is $CP$-violation 
in mixing if the parameter $|q/p| \neq 1$. The SM prediction is $A_{CP} = -(4.0 \pm 0.6)\times10^{-4}$~\cite{SM_pred_CPVMixing}.

\begin{figure}[hb!]
\begin{center}
\includegraphics[width=6.0cm,keepaspectratio]{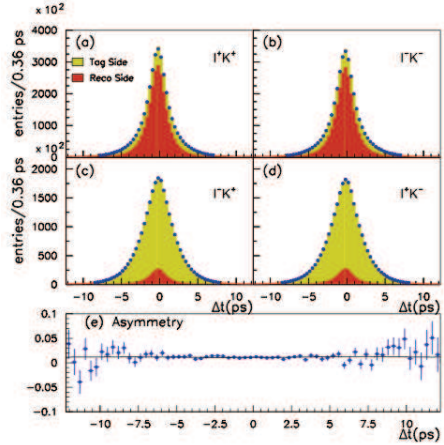}
\end{center}
\caption
{\label{fig:TRV_results}
{\em $\Delta t$ distribution for the continuum subtracted data (points with error bars) and 
fitted contribution from $K_R$ (dark) and $K_T$ (light) for $\ell^+K^+$ (top-left), $\ell^-K^-$ 
(top-right), $\ell^-K^+$ (middle-left) and $\ell^+K^-$ (middle-right) events. The bottom plot is 
the $\Delta t$-dependent raw asymmetry between $\ell^+K^+$ and $\ell^-K^-$ events.
}}
\end{figure}

The usual observable to measure the mixing $A_{CP}$ is the di-lepton asymmetry, 
$A_{CP} = \frac{N(\ell^+\ell^+) - N(\ell^-\ell^-)}{N(\ell^+\ell^+) + N(\ell^-\ell^-)}$, where $\ell = e$ or $\mu$, and $\ell^+$ 
($\ell^-$) tags a $B^0$ ($\bar{B}^0$). This measurement benefits from the high statistics but has the drawback on relying on 
control samples to subtract charge-asymmetric backgrounds. The systematic uncertainty related to this correction constitutes 
a severe limitation on the precision of the measurement. The present analysis measures $A_{CP}$ with a new technique in which 
one of the $B^0$-mesons in the event is reconstructed in $B^0\rightarrow D^{*-} X\ell^+\nu$ (referred to as the $B_R$), with 
a partial reconstruction of the $D^{*-}\rightarrow\pi^-\bar{D}^0$ decay. The flavour of the other $B^0$ (referred to as the $B_T$) 
is tagged by looking at the charge of the charged kaons in the event ($K_T$). Because a $B^0$ ($\bar{B^0}$) decays most often 
to a $K^+$ ($K^-$), then when mixing takes place $\ell$ and $K_T$ have the same charge. A kaon with the same sign as $\ell$ may 
also come from the partially reconstructed $D^0$ in the event ($K_R$). To extract $A_{CP}$, three raw asymmetries are measured,
\begin{eqnarray}
A_{\ell} &=& A_{r\ell} + A_{CP}\chi_d, \\
A_T      &=& \frac{N(\ell^+K^+_T) - N(\ell^-K^-_T)}{N(\ell^+K^+_T) + N(\ell^-K^-_T)} = A_{r\ell} + A_K + A_{CP}, \\
A_R      &=& \frac{N(\ell^+K^+_R) - N(\ell^-K^-_R)}{N(\ell^+K^+_R) + N(\ell^-K^-_R)} = A_{r\ell} + A_K + A_{CP},
\end{eqnarray}
where $A_{\ell}$ is the inclusive single lepton asymmetry, i.e. the asymmetry between events with 
$\ell^+$ compared to those with $\ell^-$, $\chi_d = 0.1862 \pm 0.0023$~\cite{chid} and $A_{r\ell}$ 
($A_K$) the detector induced charge asymmetry in the $B_R$ ($K^{\pm}$) reconstruction.

The $B_R$ is selected by combining a high momentum lepton and an opposite charge soft pion from the 
decay $D^{*-}\rightarrow\bar{D}^0\pi^-_s$, both consistent with originating from a common vertex. The $B_R$ 
events are discriminated against backgrounds by using the unobserved neutrino mass squared
${\mathcal M}^2_{\nu} = (E_{\rm beam} - E_{D^*} - E_{\ell})^2 - ({\mathbf p}_{D^*} + {\mathbf p}_{\ell})^2$, 
where the $B^0$ momentum is neglected. $E_{\ell}$ and ${\mathbf p}_{\ell}$ are the energy and momentum of 
the lepton, and ${\mathbf p}_{D^*}$ is an estimation of the of the $D^*$ momentum by approximating its 
direction the same as the $\pi^-_s$ and parameterizing its momentum as a linear function of ${\mathbf p}_{\pi^-_s}$ 
using MC. ${\mathcal M}^2_{\nu}$ peaks near zero for signal. The production point of the reconstructed $K$ 
($K$-vertex) is estimated by the intersection of its track and the beam-region. $\Delta z$ is defined as the 
distance from the $\ell\pi_s$ vertex and $K$-vertex along the beam-axis. Finally, the proper time difference 
$\Delta t$ between $B_R$ and $B_T$ is defined as $\Delta t = \Delta z/\beta\gamma$ (with $\beta\gamma = 0.56$ 
the average Lorentz boost of the $e^+e^-$ collision). The estimated error on the estimated $\Delta t$, 
$\sigma(\Delta t)$, is as well used as a discriminant variable. Events in which $\ell$ and $K$ have the 
same sign are defined as mixed and unmixed otherwise. $K_R$ candidates tend to have a smaller $\Delta t$ 
than $K_T$ candidates, therefore $\Delta t$ is used as one of the main discriminant variables. Furthermore, 
$K_R$ are usually emitted mainly back-to-back with respect to $\ell$, while $K_T$ are produced at random, so 
we use in addition the angle $\theta_{\ell K}$ between $K$ and $\ell$.

The number of $B_R$ events is extracted by fitting the ${\mathcal M}^2_{\nu}$ distributions. The events are split
in four lepton categories ($(e^{\pm},\mu^{\pm})$) and in eight tagged samples 
$(e^{\pm}K^{\pm},\mu^{\pm}K^{\pm})$ for the extraction of $A_{\ell}$ and $(A_T,A_R)$, respectively. 
A total of $(5.945 \pm 0.007)\times10^6$ peaking events are found. We 
measure $A_{CP}$ with a binned four dimensional fit to $\Delta t$, $\sigma(\Delta t)$, $\cos(\theta_{\ell K})$ 
and $p_{K}$. Figure~\ref{fig:TRV_results} show the fit projections for $\Delta t$. We find 
$A_{CP} = (0.06 \pm 0.17^{+0.38}_{-0.32})\%$, and $1-|q/p|^2 = (0.29\pm0.84^{+1.78}_{-1.61})\times10^{-3}$~\cite{B0B0Mixing_results}.
This is the single most precise measurement of this mixing asymmetry well in agreement with the SM expectations.

\section{Time-dependent amplitude analysis of $B^0\rightarrow (\rho\pi)^0$}

The $B^0\rightarrow\pi^+\pi^-\pi^0$ decay is well suited for $CP$-violation studies. The phase space 
of this final state is dominated by intermediate vector resonances ($\rho$). A complete time-dependent 
Dalitz plot (DP) analysis is sensitive to the interference between the resonant $\rho^+$, $\rho^-$ and 
$\rho^0$ intermediate states, allowing to extract the strong and weak relative 
phases, and of the $CP$-violation parameter $\alpha = {\rm arg}[-(V_{td}V^*_{tb})/(V_{ud}V^*_{ub})$, 
with $V_{qq'}$ the elements of the $CKM$ matrix~\cite{BtoRhoPi_alpha_theo}.

\begin{figure}[hb!]
\begin{center}
\includegraphics[width=3.5cm,keepaspectratio]{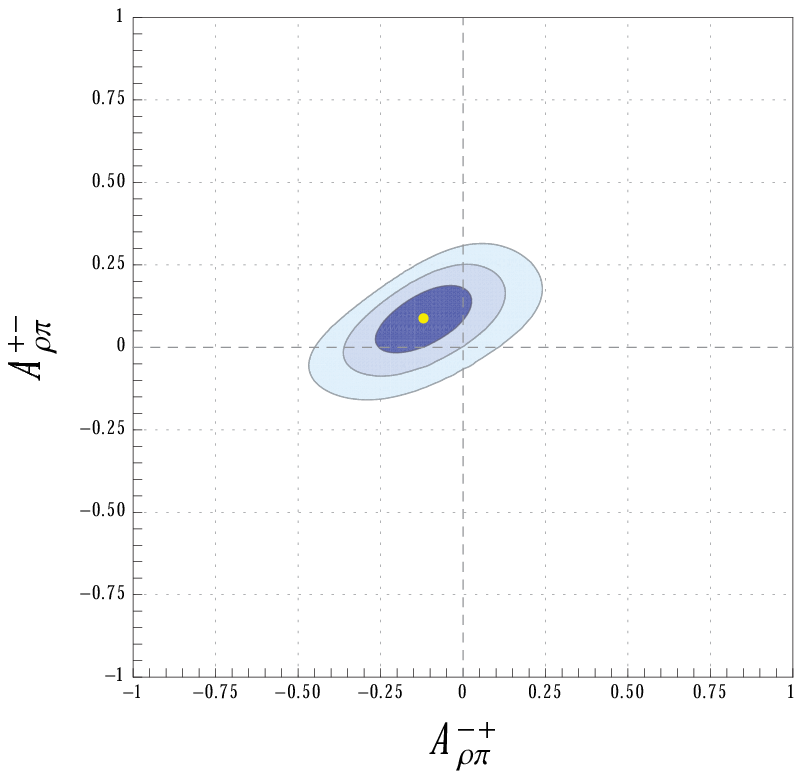}
\includegraphics[width=5.0cm,keepaspectratio]{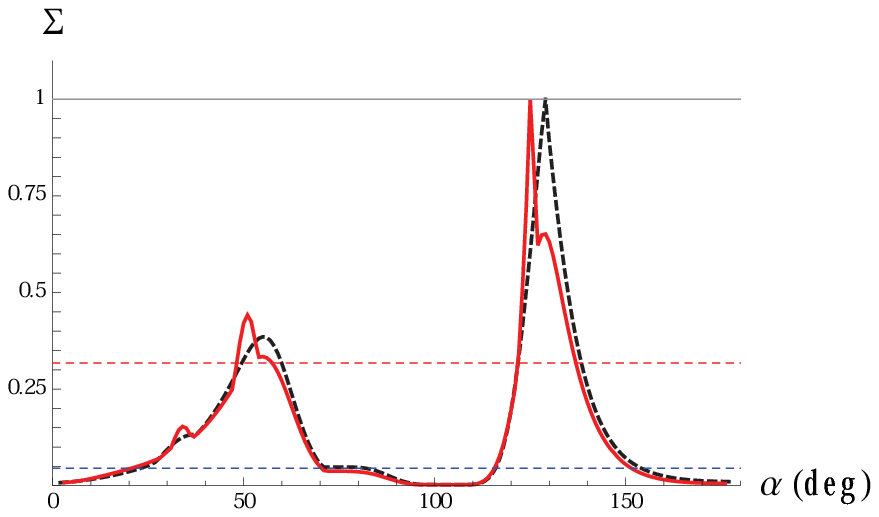}
\end{center}
\caption
{\label{fig:RhoPi_results}
{\em Left: two-dimensional likelihood scan of $A^{+-}_{\rho\pi}$ vs $A^{-+}_{\rho\pi}$
with 1,2 and 3 $\sigma$ C.L. contours. The yellow dot inside the contours indicate the central 
value. Right: Isospin-constrained (solid red) and unconstrained (dashed black) scans of $\Sigma$ 
(see text) as a function of $\alpha$.
}}
\end{figure}

The time-dependent amplitude for $B^0$ decays to the $\pi^+\pi^-\pi^0$ is given by 
$A_{3\pi} = f_+A^+ + f_-A^- + f_0A^0$, and similarly for $\bar{B}^0$ decays, with 
the $A^i$ replaced by $\bar{A}^i$ ($i=+,-,0$). The DP-dependent $f_i$, and are defined 
in terms of modified Breit-Wigner resonances. The time-dependent probability for a meson which is a 
$B^0$ ($g^-_{3\pi}$) or $\bar{B}^0$ ($g^+_{3\pi}$) at the time the other one decays, to decay to 
$\pi^+\pi^-\pi^0$ is given by,
\begin{equation}
 g^{\pm}_{3\pi}(\Delta t,DP) = \frac{e^{-|\Delta t|/\tau_{B^0}}}{4\tau_{B^0}}(|A_{3\pi}|^2 + |\bar{A}_{3\pi}|^2)
 \big(
 1
 \mp C_{3\pi}\cos(\Delta m_d\Delta t)
 \pm S_{3\pi}\sin(\Delta m_d\Delta t)
 \big)
\end{equation}
where $C_{3\pi} = \frac{|A_{3\pi}|^2 - |\bar{A}_{3\pi}|^2}{|A_{3\pi}|^2 + |\bar{A}_{3\pi}|^2}$, 
$S_{3\pi} = 2{\rm Im}\big\{\frac{(q/p)\bar{A}_{3\pi}A^*_{3\pi}}{|A_{3\pi}|^2 + |\bar{A}_{3\pi}|^2}\big\}$ 
and is $\tau_{B^0}$ the $B^0$ lifetime. 
The decay amplitudes $A_{3\pi}$ and $\bar{A}_{3\pi}$~\cite{RhoPi0_results} are written 
in terms of 27-real valued parameters $U$ and $I$ coefficients which have a number of advantages: 
there is a unique solution of the $U$-$I$ from the fit to data; their uncertainties are more Gaussian 
than those from fits where the decay amplitudes are directly parameterized in terms of the $A^i$ moduli 
and phases; and it is simpler to combine measurements from different experiments. The physical quantities 
(branching fraction, $CP$-asymmetry) for each $\rho\pi$ charge states are functions of the $U$ and $I$ 
parameters.

The present analysis~\cite{RhoPi0_results} is an update of the previous {\babar} measurement~\cite{BaBar_previous_RhoPi0} with the 
full dataset. Background events are discriminated by using two kinematic variables: 
$m^2_{ES} = [(s/2 + \vec{p}_i \cdot \vec{p}_B)/E_i]^2 - |\vec{p}_B|^2$ and 
$\Delta E = E^*_B - \sqrt{s}/2$, where $\sqrt{s}$ is the $e^+e^-$ beam energy in the CM frame, $(E_i,\vec{p}_i)$ 
and $\vec{p}_B$ the four-momentum of the $e^+e^-$ system and the momentum of the B-candidate in the laboratory 
frame, and $E^*_B$ the B-candidate energy in the CM frame. $m_{ES}$ and $\Delta E$ peak at the $B$-mass and at 
zero for signal events, respectively. Further background discrimination is achieved by using a neural-network (NN) 
which exploits the topological differences between signal and background. A maximum likelihood fit using the 
$\Delta t$ and DP variables, as well as $m_{ES}$, $\Delta E$ and NN, is performed to extract the values of the $U$-$I$ coefficients. 
Two direct $CP$-violation parameters,
\begin{equation}
A^{+-}_{\rho\pi} = \frac{\Gamma(\bar{B}^0\rightarrow\rho^-\pi+) - \Gamma(B^0\rightarrow\rho^+\pi-)}{\Gamma(\bar{B}^0\rightarrow\rho^-\pi+) + \Gamma(B^0\rightarrow\rho^+\pi-)}
,~~~
A^{-+}_{\rho\pi} = \frac{\Gamma(\bar{B}^0\rightarrow\rho^+\pi-) - \Gamma(B^0\rightarrow\rho^-\pi+)}{\Gamma(\bar{B}^0\rightarrow\rho^+\pi-) + \Gamma(B^0\rightarrow\rho^-\pi+)}
\end{equation}
are extracted with the values $A^{+-}_{\rho\pi} = 0.09^{+0.05}_{-0.06}\pm 0.04$ and $A^{-+}_{\rho\pi} = -0.12\pm0.08^{+0.04}_{-0.05}$. 
A two-dimensional likelihood scan is provided in the left hand plot of figure~\ref{fig:RhoPi_results}. The origin, corresponding 
to no direct $CP$-violation, is excluded at the level of $\sim 2\sigma$.

Scans of the likelihood function in fits where a given value of the $CKM$ $\alpha$ angle is assumed are performed enforcing the 
$SU(2)$ symmetry in a loose ({\it unconstrained} analysis using only the $B^0\rightarrow\rho\pi$ amplitudes) or tight 
({\it constrained} analysis adding the charged $B^+\rightarrow \rho \pi$ amplitudes) fashion.
The $\Sigma$ scan vs $\alpha$ is shown in the right hand 
plot of figure~\ref{fig:RhoPi_results}. The $\Sigma$ value is commonly referred as "1 - C.L.", however robustness studies 
have shown that with the current data sample the $\Sigma$ cannot interpreted in terms of the usual Gaussian 
statistics~\cite{RhoPi0_results}. Hence with the current statistics, the analysis cannot reliably determine 
the angle $\alpha$. This analysis would benefit greatly from increased sample sizes available 
at higher-luminosity experiments.

\section{Conclusion}
We presented the fist direct observation of $T$-reversal violation in the B-meson system, which is 
established at the level of $14\sigma$. Deviations of $CPT$ conservation are also tested giving null 
results, in agreement with the expectations of the $CPT$-theorem. We also reported on a new experimental 
technique for the measurement of the mixing induced $CP$-violation parameter $1 - |q/p|^2$. The measurement 
is the most precise single measurement up to date and is well in agreement with the SM expectations. 
Finally, we reported on the update using the full {\babar} dataset of the time-dependent amplitude analysis 
of the $B^0\rightarrow\pi^+\pi^-\pi^0$ decays. Measurements of direct $CP$-violation asymmetries are measured,
excluding the $CP$-violation conservation hypothesis at the level of $~2\sigma$. Constrains on the $\alpha$ 
CKM angle are calculated. Robustness studies show that with the current dataset the method for extracting 
$\alpha$ is not robust, meaning that the current constrains cannot be interpreted in terms of the usual 
Gaussian statistics, but the analysis should benefit from increased data-samples from future experiments.


\begin{thebibliography}{99}
\bibitem{CKM} N. Cabibbo, {\it Phys. Rev. Lett.} {\bf 10}, 531 (1963);
M. Kobayashi and T. Maskawa, {\it Prog. Theor. Phys.} {\bf 49}, 652 (1973).
\bibitem{CPV_K} J. H. Christenson {\it et al.} {\it Phys. Rev. Lett.} {\bf 13}, 138 (1964).
\bibitem{CPV_B} B. Aubert {\it et al.} ({\babar} Collaboration), {\it Phys. Rev. Lett.} {\bf 87}, 091801 (2001); 
K. Abe {\it et al.} (Belle Collaboration), {\it Phys. Rev. Lett.} {\bf 87}, 091802 (2001);
B. Aubert {\it et al.} ({\babar} Collaboration), {\it Phys. Rev. Lett.} {\bf 93}, 131801 (2004);
Y. Chao {\it et al.} (Belle Collaboration), {\it Phys. Rev. Lett.} {\bf 93}, 191802 (2004).

\bibitem{CPT} G. L\"uders {\it et al.} 1954, Vol. 28, p. 5;
J. S. Bell, Ph.D. thesis, Birmingham University, 1954;
Niels Bohr and the Development of Physics, edited by W. Pauli, L. Rosenfold, and V.
Weisskopf (McGraw-Hill, New York, 1955).
\bibitem{CPT_exp} R. Carosi {\it et al.}, {\it Phys. Lett.} B {\bf 237}, 303 (1990);
A. AlaviHarati {\it et al.}, {\it Phys. Rev.} D {\bf 67}, 012005 (2003);
B. Schwingenheuer {\it et al.}, {\it Phys. Rev. Lett.} {\bf 74}, 4376 (1995).
\bibitem{TRV_K} A. Angelopoulus {\it et al.} (CPLEAR Collaboration), {\it Phys. Lett.} B {\bf 444}, 43 (1998).

\bibitem{TRV_method} J. Bernabeu {\it et al.}, {\it J. High Energy Phys.} {\bf 08} (2012) 064.
\bibitem{TRV_paper} J. P. Lees {\it et al.}, {\it Phys. Rev. Lett.} {\bf 109}, 211801 (2012).

\bibitem{CPV_types} G. C. Branco, L. Lavoura and J. P. Silva, {\it CP Violation},
International Series of Monographs on Physics (1999).

\bibitem{SM_pred_CPVMixing} A. Lenz, U. Nierste, arXiv:1102.4274 [hep-ph] (2010);
U. Nierste, arXiv:1212.5805 [hep-ph] (2012);
J. Charles {\it et al.}, {\it Phys. Rev.} D {\bf 84}, 033005 (2011).
\bibitem{chid} J. Beringer et al., (Particle Data Group), {\it Phys. Rev.} D {\bf 86}, 010001 (2012).
\bibitem{B0B0Mixing_results} J. P. Lees {\it et al.}, arXiv:1305.1575 [hep-ex] (2013).

\bibitem{BaBar_previous_RhoPi0} B. Aubert {\it et al.} ({\babar} Collaboration), {\it Phys. Rev.} D{\bf 76}, 012004 (2007).
\bibitem{BtoRhoPi_alpha_theo} H.R. Quinn and A.E. Snyder, {\it Phys. Rev.} D {\bf 48}, 2139 (1993).
\bibitem{RhoPi0_results} J. P. Lees {\it et al.}, arXiv:1304.3503 [hep-ex] (2013).

\end{thebibliography}
\end{document}